
\input harvmac
\Title{FERMI-PUB-92/110-T}{Asymptotic Behavior of 2-d Black Holes}
\centerline{{\bf E. Raiten}\footnote{$^\dagger$}
{e-mail:\ Raiten@FNAL}}
\bigskip\centerline{Theory Group, MS106}
\centerline{Fermi National Accelerator Laboratory}
\centerline{P.O. Box 500, Batavia, IL 60510}

\vskip .3in
We consider solutions of the field equations for the large $N$ dilaton
gravity model in $1+1$ dimensions of Callan, Giddings, Harvey,
and Strominger (CGHS).  We find time dependant solutions in the weak
coupling region with finite mass and vanishing flux, as well as
solutions with lie entirely in the Liouville region.
\Date{4/92}

%
\def\cmp#1{{\it Comm. Math. Phys.} {\bf #1}}

\def\prd#1{{\it Phys. Rev.}{\bf D#1}}

\def\sp{\varphi}
\def\r{\rho}
\def\p{\phi}
\def\dpr{\partial_+\rho}
\def\dmr{\partial_-\rho}
\def\dpp{\partial_+\phi}
\def\dmp{\partial_-\phi}
\def\er{e^{2\rho}}
\def\emr{e^{-2\rho}}
\def\s{\sigma}

\def\emp{e^{-2\phi}}
\def\na{{N\over {12}}}
\def\nb{{N\over {24}}}
\def\t{\tau}
\def\l{\lambda}
\def\d{\delta}
\def\nc{{N\over {48}}}
\def\es{e^{2\l\s}}
\def\ems{e^{-2\l\s}}
\def\ml{{M\over {2\l}}}

\def\ppm{\partial_+}
\def\pp{\partial_-}
\def\ps{\psi}
\nref\hawa{S. Hawking, \cmp{43} (1975) 199.}
\nref\cala{C. Callan, S. Giddings, J. Harvey, A. Strominger, \lq
\lq Evanescent Black Holes," UCSB-TH-91-54, 11/91.}
\nref\bir{B. Birnir, S. Giddings, J. Harvey, A. Strominger, \lq\lq
Quantum Black Holes, UCSB-TH-92-08, 3/92.}
\nref\hawb{S. Hawking, \lq\lq Evaporation of Two Dimensional Black Holes,"
CALT-68-1774, 2/92.}
\nref\russ{J. Russo, L. Susskind, L. Thorlacius, \lq\lq Black Hole
Evaporation in 1+1 Dimensions," SU-ITP-92-4, 1/92.}
\nref\suss{L. Susskind, L. Thorlacius, \lq\lq Hawking Radiation and Back-
Reaction," SU-ITP-92-12, 3/92.}
\nref\bana{T. Banks, A. Dabholkar, M. Douglas, M. O'Loughlin, \lq\lq
Are Horned Particles teh Climax of Hawking Evaporation," RU-91-54.}
\nref\gid{S. Giddings, A. Strominger, \lq\lq Dynamics of Extremal Black Holes,"
UCSB-TH-92-01, to appear in {\it Phys. Rev.}{\bf D}.}
\nref\wit{E. Witten, \prd{44} (1991), 314.}
\nref\gidb{S. Giddings, \lq\lq Black Holes and Massive Remnants,"
UCSBTH-92-09, 3/92.}
\nref\col{S. Coleman, J. Preskill, F. Wilczek, \lq\lq Quantum Hair on
Black Holes"Black Holes," IASSNS-HEP-91/64, 12/91.}
\nref\deal{S.P. DeAlwis, \lq\lq Comments on No-Hair Theorems and Stability
of Blackholes," COLO-HEP-267, 1/92.}
\nref\pres{J. Preskill, P. Schwarz, A. Shapere, S. Trivedi, F. Wilczek,
Mod. Phys. Lett. {\bf A6} (1991) 2353.}
\vfill
\eject

\newsec{Introduction}
\medskip

In the years following the discovery of Hawking radiation and the associated
evaporation of black holes \hawa , there have been many efforts to
either prove or refute the resulting implication that an initially pure
state can collapse into a black hole and evaporate into a mixed state.
The fact that such efforts have not proven successful is due to a
combination of complications, including principally those of the
backreaction of the Hawking radiation on the metric, and of the regions of
large curvature (and hence strongly coupled quantum gravity effects) which
are expected in gravitational collapse.

Recently, Callan et. al. (CGHS) proposed a model which seemed to avoid some
of these difficulties \cala .  It consistts of gravity coupled to a dilaton
and conformal matter in $1+1$ dimensions.  For a single matter field it was
found the backscatter (i.e., the Hawking radiation) occured in a region
of strong coupling.  By proliferating the number N of matter fields, it
was believed that the essential physics was occur in a region of small
coupling and hence be amenable to a systematic $1/N$ semiclassical
expansion.

These initial hopes were dashed \bana \russ\ by the observation
that the dilaton develops a singlurity at a finite value, dependent on
$N$, precisely in the region where quantum fluctuations begin to
become large.  As a result, a number of groups \hawb \suss
\bir\ have recently tried to explore, both numerically and
analytically, the solutions of the large $N$ field equations.  In
particular, one is interested in the final \lq\lq endpoint" of the
Hawking radiation.  Therefore, in \hawb \suss \bir the
fields were assumed to depend only on a \lq\lq spatial" coordinate
(of which there are a few natural choices).  For example, in \bir ,
a series of solutions with finite ADM mass and vanishing incoming and
outgoing flux were found.  Starting at weak coupling at spatial infinity,
they were found to \lq\lq bounce" back to weak coupling in the region
of the singularity mentioned above.

The static approximation used to derive these results is a significant
simplification, but makes it difficult to consider the approach to
the endpoint of the Hawking process.  In the following, we will consider
time-dependent (approximate) solutions to the CGHS equations.  We will
find solutions which still have finite ADM mass and vanishing flux, as
well as regions with a time-dependent singular event horizon.
In a later section, we will also
discuss a series of perturbative, time dependant solutions which
lie entirely in the Liouville region, followed by some concluding
remarks.

\medskip
\newsec{The CGHS Model}
The CGHS model of dilaton gravity coupled to $N$ conformal matter fields
in $1+1$ dimensions with coordinates $\s$ and
$\t$ is defined by the action
\eqn\acta{S={1\over {2\pi}}\int d^2\sigma \sqrt{-g}[\emp (R+4(\del \p )^2
+4\lambda ^2)-{1\over 2}\sum_{i=1}^N(\del f_i)^2],}
where $g$,$\p$, and $f_i$ represent the metric, dilaton, and matter fields,
respectively, and $\lambda^2$ is the cosmological constant.  Integrating
out the matter fields and going to conformal gauge, where
\eqn\conf{g_{+-}=-{1\over 2}\er,}
\eqn\confb{g_{++}=g_{--}=0,}
($x_{\pm}=\t \pm \s$), the resulting action is
\eqn\actb{S={1\over {\pi}}\int d^2\s [\emp (\partial_+(2\p -\r )\partial_-
(2\p -\r )-\lambda^2\er )+(\na -\emp )\dpr\dmr ].}
The equations of motion for $\r$ and $\p$ are
\eqn\tpm{T_{+-}=\emp(2\partial_+\dmp -4\dpp\dmp -\lambda^2\er )-\na
\partial_+\dmr =0,}
\eqn\dil{-4\partial_+\dmp+4\dpp\dmp+2\partial_+\dmr +\lambda^2\er =0.}
Since the gauge has been fixed as in \confb , there are two constraint
conditions, namely,
\eqn\tpp{T_{\pm\pm}=\emp (4\partial_{\pm}\p\partial_{\pm}\r
-2\partial_{\pm}^2\p )-\na (\partial_{\pm}\r\partial_{\pm}
-\partial_{\pm}^2\r+t_{\pm}(\sigma_{\pm}))=0,}
where the functions $t_{\pm}$ are fixed by boundary conditions.

The simplest and most important nontrivial solution of \tpm and \dil
is the linear dilaton vacuum
\eqn\lin{\r=0,\ \p=-{{\lambda}\over 2}(\sigma^+-\sigma^-).}
This vacuum has a singulrity at
\eqn\sing{\p =\p_{cr}=-{1\over 2}ln{\na},}
as seen by calculating the sign of the kinetic operator in \actb .
As in previous papers, we will call the region of $\p < \p_{cr}$ the
dilaton region, and $\p > \p_{cr}$ the Liouville, or strong
coupling region.

\medskip
\newsec{Finite Mass Solutions}

Solutions to \tpm\ and \dil\ with finite ADM mass were first found in
\bir by assuming that both $\p$ and $\r$ are time independant.
In that case, \tpm\ and \dil\ become
\eqn\tpms{T_{+-}=\emp(-{1\over 2}\p{''}+\p{'2}-\lambda^2\er )+
{N\over {48}}\r{''}=0,}
\eqn\dils{\p{''}-\p{'^2}-{1\over 2}\r{''}+\lambda^2\er ,}
where the primes denote $d/d\s$.  Linearizing about the linear
dilaton vacuum solution \lin , for vanishing incoming and outgoing
flux $t_{\pm}$,
asymptotically the resulting equations can be expressed as
\eqn\lins{2\l \delta\p '+2\l^2\delta\r-\l\delta\r '=0,}
\eqn\lindb{\delta\p{''}=\delta\r{''}(1-\nb\ems ).}
The asymptotic form of the solutions of these equations is
\eqn\asys{\delta\p=-\ml\ems+\dots ,}
\eqn\asysb{\delta\r=-\ml\ems+\dots ,}
where the parameter $M$ is the ADM mass, given by evaluating
\eqn\adm{M=2\es (\l\delta\r+\delta\p ').}
at spatial infinity.

Before going beyond the static case, it should be noted that one can
expand $\d\p$ and $\d\l$ in powers of $\epsilon =\ems$,
\eqn\expan{\d\p =\sum_{n=1}^{\infty}a_n\epsilon^n,\
\d\r =\sum_{n=1}^{\infty}b_n\epsilon^n,}
with $a_1=b_1=-\ml$.
Substituing into the full linearized equations, one finds the relations
\eqn\rel{(1-n^2)a_{n+1}-b_{n+1}+\nb n^2b_n=0,}
\eqn\relb{(n+1)b_n=2n a_n,}
from which one easily finds
\eqn\relc{a_{n+1}=\na{{n^3}\over {n^3+n^2-n-1}}a_n.}
For large $n$, this suggests that we must have $\na\leq 1$
for the series to
converge.  For example, for $\na =1$, the resulting series
for $\d\p$ is roughly
\eqn\rough{\d\p\sim {\ml}{1\over {\es -1}},}
thus implying that our linearizing
approximation is breaking down for small $\s$.
It is perhaps of interest that the requirement $\na\leq 1$ implies
\sing\ that $\p_{cr}>0$
so that the effective critical coupling constant $e^{2\p_{cr}} >1$.

Let us now proceed beyond the static limit, but continue to require
a finite ADM mass.
Including time derivatives, in the $\s$, $\t$ coordinate
system the linearized equations read
\eqn\lina{\er (-{1\over 2}\d\p ''-2\l\d\p '-2\l^2\d\r +{1\over 2}
\d\ddot\p )+\nc (\d\r ''-\d\ddot\r )=0,}
\eqn\linb{\d\p ''-\d\ddot\p+2\l\d\p ' +{1\over 2}\d\ddot\r-{1\over 2}
\d\r '' +2\l^2\d\r =0.}
{}From \adm , we see that finite ADM mass requires
both $\delta \r$ and $\delta\p$ vary asymptotically as $\ems$, as in
\asys and \asysb .  If we express the perturbations about the linear
dilaton vacuum as
\eqn\dphi{\delta\p =x(\t )\ems ,}
\eqn\drho{\delta\r =y(\t )\ems ,}
then to leading order \lina\ and \linb\ become, respectively,
\eqn\lint{2\l^2 x-2\l^2y+{1\over 2}\ddot x=0,}
\eqn\lind{{1\over 2}\ddot y-\ddot x=0.}
It is a simple matter now to assume that $x$ and $y$ both vary
as $e^{\omega \t}$ and solve for $\omega$ and the relative amplitudes.
Of course, one solution is just
\eqn\onesol{x=y=\ml +a\t}
as in \asys (where $a=0$).
The other solution is easily seen to be
\eqn\solb{x(\t )={1\over 2}y(\t )=ae^{-2\l\t}+be^{2\l\t}.}
Substituting \solb into \adm , we seen that the time dependancy of
$\delta \p$ and of $\delta \r$ cancel, and the ADM mass is constant,
even thought the metric and the dilaton are certainly not.  Presumably,
we should set the coefficient $b$ in $\solb$ to zero, so that the
solution is well behaved as $\t\rightarrow\infty$, as should the
coefficient of the linear term in the $\omega =0$ solution.

The behavior of these solutions can be understood in much the same manner
as in the static case \bir .  Let us concentrate on the $M=0$ solution,
as it has been suggested that it represents the true quantum vacuum of
the theory \bir .  In any case, for $\t$ sufficiently negative, the
time dependant terms dominate over the static terms.  As one integrates
the equation of motion in from spatial infinity, the solution may approach
the singularity at $\p_{cr}$ (in the static case, this approach was
guaranteed).  In this region, we can essentially set $\r =0$, and
$\p =\p_{cr}+\sp$.  The resulting equation of motion is
\eqn\bnca{-\sp (\sp ''-\ddot\sp)={1\over 2}(\sp '^2-\dot\sp^2-\l^2).}
If we continue to assume that $\dot\sp=-2\l\sp$, then \bnca can be
integrated, yielding
\eqn\bncb{\sp '^2-{A\over {\sp}}-{{10}\over 3}\l^2\sp^2=\l^2,}
where $A$ is an integration constant.   As long as $A\neq 0$, this is
the equation for a particle in a potential with an infinite barrier at
the origin, so $\sp$ will bounce back to the weak coupling regime.

We can also discuss the behavior of the solutions
for any region where $\r\rightarrow -\infty$,
in particular as $\s\rightarrow
-\infty$, assuming that $ae^{-2\l\t}<\ml$, as was discussed in the
static case in \bir , by dropping terms proportional to $\er$ which
become irrelevant for $\r\rightarrow -\infty$.
For in that case we
have
\eqn\infa{\emp =-\na \r+a_+\s_++a_-\s_-+b,}
\eqn\infb{e^{-\p}\sqrt{\emp-\na}-\na ln[\sqrt{\emp-\na}+e^{-\p}]=
f(\s_-)+g(\s_+),}
where $a_{\pm}$ and $b$ are constants,
and $f$ and $g$ are arbitrary functions
of their arguments (in the static case \bir , one has $f+g=-a\s +c$),
the only priviso being that $f$ must be smooth (i.e.,
$f(\s_-)$ is the integral of a completely arbitrary function).
Concentrating on a region where $f+g\rightarrow\infty$, we have
\eqn\sina{\emp\sim f+g+\nb ln(f+g),}
\eqn\sinb{\r\sim -{{12}\over N}(f+g)-{1\over 2}ln (f+g)-a_+\s_+-a_-\s_--b.}
Using the formula for the curvature,
\eqn\curv{R=8\emr\partial_+\dmr ,}
we have
\eqn\sinc{R\sim {{\partial_+g\partial_-f}\over {f+g}}e^{{{24}\over N}
(f+g-a_+\s_+-a_-\s_--b)},}
(where we have redefined the constants $a_{\pm}$ and $b$).
Taking, for example, $g(\s_+)\sim (\s_+ -\s_+^0)^{-\alpha}, \alpha >0$,
we see that $\s_+^0$ is a singular event horizon.  Since $\s_+=\t+\s$,
the location of the horizon is not constant in time $\t$.
Furthermore, the fact that $\d\r$ grows more rapidly than $\d\p$, as
seen in \solb , suggests that such regions might be of greater importance
in understanding the full evolution of the system, particularly for the
$M=0$ solution, which has been proposed to be the true vacuum of the
theory.
In fact, in the original, unperturbed field equation \tpm\ , we see
that if $\p '=\dot\p =-2\l\p$, then $\r$ is forced to approach $-\infty$,
unless $e^{2\p}\sim 24/N$.  Of course, at this point,
depending on $N$, we
may no longer be in the weak coupling regime which we have been discussing,
but rather in the strong coupling, or Liouville region, which we
consider below.

Of course, for large $\t$, the time dependant terms are small, and the
solution behaves as in the static case, where $\p$ penetrates closer
and closer to $\p_{cr}$ before bouncing back to weak coupling \bir .
But for $\t$ sufficiently large and negative, we are effectively
dealing with the $M=0$ solution, in which $\r$ will tend to grow
faster than $\p$ and singular event horizons should appear.  It is
questionable whether or not this is a reasonable condition for the
true vacuum of the theory.  Actually, it seems more reasonable that
the final state of the system, in response to some incoming matter,
would have a potentially complicated causal structure.  Of course, our
solutions are nonsingular at $\p_{cr}$ whereas the incoming matter is
singular there, so the interpretation of these solutions remains unclear.

To complement these solutions, we should in prinicple search for time
dependant solutions with regular horizons, as was done in \suss , \bir ,
and \hawb , generally by using the \lq\lq spatial" variable $s=x_+x_-$
and then imposing continuity conditions at the horizon at $s=0$.
Including time dependant terms, of course, will affect the location of
the horizon in general, and we have not yet made a determined effort
to analyze the range of possibilities.  Work on this problem is in
progress.

\medskip
\newsec{The Liouville Region}

As argued in \bana , \gid , solutions which lie entirely in the Liouville
region contain important information concerning the behavior of extremal
four-dimensional dilaton black holes. Secondly, it might be possible that
a configuration in the Liouville region might evolve into the weak coupling
region, even if the reverse is impossible.

To analyze this region, we introduce the new dependant variable \bir
\eqn\liou{\psi =e^{-\p},}
in terms of which the action is just
\eqn\actlio{S={1\over {\pi}}\int d^2\s (4\partial_+\psi\partial_-\psi
+4\psi\partial_+\psi\partial_-\r-\l^2\psi^2\er +\na\dpr\dmr ).}
The resulting field equations (which can just as easily be derived from
the original field equations upon substituting \liou\ ) are
\eqn\lflda{T_{+-}=-2\partial_+\psi\partial_-\psi
-2\psi\partial_+\partial_-\psi-\l^2\psi^2\er-\na\partial_+\partial_-\r=0,}
\eqn\lfldb{4\partial_+\partial_-\psi+2\psi\partial_+\partial_-\r +
\l^2\psi\er =0.}
The simplest solution to these equations is the trivial solution
\eqn\lsola{\psi =0, \r=0.}
If we now perturb these equations about \lsola , we see that every
term in \lflda\ is quadratic except the last term, so we just have
\eqn\free{\d\r =f_+(\s_+)+f_-(\s_-),}
where $f_{\pm}$ are arbitraty functions.  Similarly,
the linearization of \lfldb\ yields simply the Klein Gordon equation
\eqn\kg{\partial_+\partial_-\d\psi+{{\l^2}\over 4}\d\psi=0}
for a particle with $m^2=\l^2/4$.

Another solution of \lflda\ \lfldb\ is \bir
\eqn\lsolb{\psi^2=\nb ,}
\eqn\lsolbb{\r =-{\rm ln}(\sqrt{2}\l\s ),}
which is an example of anti-deSitter space, as the curvature turns out to
be $R=-4\l^2$.
Linearizing again, we find
\eqn\la{-2\sqrt{\nb}\pp\ppm\d \ps -\na\pp\ppm\d \ps-\nb {1\over {\s^2}}\d\r
 -\sqrt{\nb}{{\d\ps}\over {\s^2}}=0,}
and
\eqn\lb{4\sqrt{\nb}\pp\ppm\d\ps +\na\pp\ppm\d\r +\nb{1\over {\s^2}}\d\r =0.}
Adding the equations, we have
\eqn\lc{2\pp\ppm\d\ps-{{\d\ps}\over {\s^2}}=0,}
which is just the equation for a particle in a $1/r^2$ potential.
For example, going to the static limit, we have
\eqn\geoa{\d\ps ''=-{2\over {\s^2}}\d\ps ,}
with solutions
\eqn\geob{\d\ps =a_1\s^{\beta_1}+a_2\s^{\beta_2},}
where the $a_i$ are constants and the $\beta_i$ are the solutions of
the quadratic equation $x^2-x+2=0$.  Since the $\beta_i$ are
therefore complex, whereas $\ps$ should be real, it would seem that
this is an inappropriate background for such a perturbative analysis.

\medskip
\newsec{Discussion}

Spurred on in part by recent advances in string theory \wit\ , we
have witnessed a great increase in the number of toy models,
particularly in low dimensions, made available for the study of
phenomena such as Hawking radiation
and the final state of black holes which involve fundamental
issues surrounding quantum gravity.  The CGHS model is an especially
simple yet sufficiently rich example of such a model.  Unfortunately,
there remain significant barriers which interfere with our
greater understanding of quantum gravity.  Of the various groups who
have studied the CGHS system, there are adherents of a variety of
scenarios, including naked singularities \hawb\ , macroscopic objects
\gidb\ , the \lq\lq bounce" scenario \bir\ , and so on.

In this letter, we have tried to begin the program of going beyond the
static limit applied earlier \bir\ \hawb\ \suss .  We know that the
classical no-hair theorems, which essentially say that a black hole
is characterized by the quantum numbers of long range fields, such as
mass, charge, angular momentum, cannot contain quantum
mechanical information.  What we have found is that specifying the
mass of the black hole does not fully specify the metric or
dilaton, even to leading order asymptotically.  There is
active research underway on
a variety of quantum-mechanical effects on black holes, see \col
for example for a thorough discussion of quantum hair and
Aharonov-Bohm type interactions of black holes.

In the present case, in the original CGHS model (i.e.,
$N=1$), the picture of the black hole was of an asymptotically flat
plane connected via a throat-like horizon to a semi-infinite
cylinder-like region.  When matter impinges on this system,
one might imagine, for example, that while
the asymptotically flat region would eventually see a constant
mass, the matter might be hurtling down the cylinder
behind the event horizon in a complicated
and possibly singular fashion.  Even the horizon itself need not be
fixed, though of course that would be measureable to an asymptotic
observer.

Another important factor which we have come across is the problem
of the crossover between weak coupling and Liouville regions.  In
spite of the initial hopes, it appears that the important physics
is occuring precisely in this region, where we cannot ignore futher
quantum corrections.  This region is small (of order $\l^{-1}$) in
the large $N$ limit, so the model may yet be viable for questions
regarding longer range phenomena.  Furthermore, because of this
great uncertainty, we cannot say for certain that propagation
through the apparant singularity is in fact forbidden.  Perhaps
a further exploration of the appropriate boundary conditions or
additional terms in the $1/N$ expansion
will suggest a way out of our present dilemnas.

\medskip
\noindent{{\bf Acknowledgements}: The author would like to thank J. Lykken,
S. Chaudhuri, H. Dykstra and J.D. Cohn for useful discussions.}

\vfill
\eject
\listrefs
\bye